\documentclass[preprint,authoryear,12pt]{elsarticle}

 \usepackage{graphicx}

\usepackage{amssymb}
\usepackage{amsfonts}
\usepackage{amsbsy}
\usepackage{amsmath}

\bibpunct{(}{)}{;}{a}{}{,}

\newcommand{\rb}{\boldsymbol{r}}
\newcommand{\Rb}{\boldsymbol{R}}

\newcommand{\ph}{\widehat p}

\newcommand{\xh}{\widehat x}

\newcommand{\Mmatc}{\mathcal{M}}

\def \xmm {XMM-Newton}

\def \approxgt{\mathrel{\hbox{\rlap{\lower.55ex \hbox {$\sim$}}
       \kern-.3em \raise.4ex \hbox{$>$}}}}
\def \approxlt{\mathrel{\hbox{\rlap{\lower.55ex \hbox {$\sim$}}
       \kern-.3em \raise.4ex \hbox{$<$}}}}

\def \exo {EXO\,0748$-$676}

\journal{Astronomy and Computing}

\begin{document}

\begin{frontmatter}



\title{Irregular time series in astronomy and the use of the Lomb-Scargle periodogram}

\author[1]{R. Vio}
\author[2]{M. Diaz-Trigo}
\author[2,3]{P. Andreani}

\address[1]{Chip Computers Consulting s.r.l., Viale Don L.~Sturzo 82,
              S.Liberale di Marcon, 30020 Venice, Italy}
\address[2]{ESO, Karl Schwarzschild strasse 2, 85748 Garching, Germany}
\address[3]{INAF-Osservatorio Astronomico di Trieste, via Tiepolo 11, 34143 Trieste, Italy}

\begin{abstract}
Detection of a signal hidden by noise within a time series is an important problem in many astronomical searches, i.e. for light curves containing the contributions of periodic/semi-periodic 
components due to rotating objects and all other astrophysical time-dependent phenomena. One of the most popular tools for use in such studies is the {{\it periodogram}},  whose use
in an astronomical context is often not trivial. The {{\it optimal}} statistical properties of the periodogram are lost in the case of irregular sampling of signals, which is a common situation in  astronomical experiments.
Parts of these properties are recovered by the {{\it Lomb-Scargle}} (LS) technique, but at the  price of theoretical difficulties, that can 
make its use unclear, and of algorithms that require the development of dedicated software if a fast implementation is necessary.
Such problems would be irrelevant if the LS periodogram could be used to significantly improve the results obtained by approximated but
simpler techniques. In this work we show that in many astronomical applications simpler techniques provide
results similar to those obtainable with the LS periodogram. The meaning of the {{\it Nyquist frequency}} is also discussed in the case of irregular sampling.
\end{abstract}


\end{frontmatter}



\section{Introduction}
The search for characteristic frequencies in astrophysical phenomena requires a careful analysis of the data with appropriate statistical tools.
Given the simplicity of its use and the wide availability of efficient related software, one of the most popular techniques for looking for periodicities within
a time series is the periodogram technique. In  astronomical applications, however, the use of this technique is not trivial.
In fact, this tool exhibits its {\it optimal} properties only in the case of signals sampled on a regular time grid, a common situation in engineering applications but not always in astronomical experiments. 
The analysis of a periodogram in the case of irregular sampling is often limited by the possibility for fully fixing its statistical properties. This is an old dated problem \citep[e.g.][]{got75} and there have been many attempts to solve it.
A partial solution has been found in the {{\it Lomb-Scargle}} (LS) approach \citep{lom76, sca82} but at price of theoretical difficulties that make its use unclear
and, if a fast implementation is needed (e.g. in the case of very long time series),  the necessity of dedicated software. Of course,
this would not constitute a relevant issue if LS periodogram could be used to notably improve the results
obtainable by the statistical analysis of a time series.
In this paper we argue that in astronomical applications often this is not the case.
We show how the negligible improvements obtained with LS are offset by the ease of interpretation and clarity  of the results provided by simpler techniques, which
do not demand high computing power and/or complicated algorithms.

In Sec.~\ref{sec:regular} the statistical analysis of sampled signals is addressed in the case of a regular sampling, where the mathematical notation and formalism are also outlined. The problems and  advantages
of an irregular sampling are analyzed in Sec.~\ref{sec:irregular}. The real advantage  of the LS periodogram with respect  to an approximated but simpler technique is considered in
Sec.~\ref{sec:spectrogram} on the 
basis of theoretical arguments as well as numerical experiments based on synthetic data and  of an experimental time series. Finally, Sec.~\ref{sec:conclusions} derives our conclusions.

\section{Statistical analysis of regularly sampled signals}  \label{sec:regular}
If  a signal $x(t)$  is sampled on a regular time grid with a constant time step $\Delta t$, a time series  $\{ x_j \}_{j=0}^{N-1} \equiv (x_{0}, x_{1}, \ldots, x_{N-1} )$  is obtained\footnote{Typically it is  assumed that $\Delta t = 1$.}.
Often the main problem is testing whether $x(t)$ is due only to a noise $n(t)$,  or whether some other
component $s(t)$ is present, i.e.  $x_j = s_j + n_j$. The most popular approach consists of computing the {\it periodogram} $\{ p_k \}_{k=0}^{N-1}$ for a set of $N$ equispaced frequencies $\{ f_k \}_{k=0}^{N-1} \equiv \{k/N\}$: $p_k =  \frac{1}{N} \left| \xh_k  \right|^2$
with the  {\it discrete Fourier transform} (DFT) of $\{ x_j \}$ being
\begin{equation} \label{eq:dft}
\xh_k = \sum_{j=0}^{N-1} x_j {\rm e}^{-i 2 \pi k j / N}, \quad k=0, 1, \dots, N-1;
\end{equation}
and $\{ f_k \}$ being the {\it Fourier frequencies}.  
The original time series $\{ x_j \}$ can be recovered from $\{ \xh_k \}$ via
\begin{equation} \label{eq:dfti}
x_j = \frac{1}{N} \sum_{k=0}^{N-1} \xh_k {\rm e}^{i 2 \pi k j / N}, \quad j=0, 1, \dots, N-1.
\end{equation}
In the case where $\{ x_j \}$ is only noise with  $\{ n_j \}$ a zero-mean, Gaussian, white-noise stationary process with standard deviation $\sigma_n$, from Eq.~(\ref{eq:dft}) 
it can be readily verified that, independently of $k$, $\ph_k/\sigma_{n}^2$ is given by the sum of two  squared independent, zero-mean, unit-variance, Gaussian random quantities. 
As a consequence, the corresponding {\it probability density function} (PDF) is the exponential distribution.  Moreover,  whenever $k \ne k'$ with $k,k'=0, 1, \ldots, N/2$, $p_k$ is independent of 
$p_{k'}$. Hence, the probability $\alpha$ that at least one of the $p_k$ is expected to exceed a level $L_{{\rm Fa}}$ is
\begin{equation}
\alpha = 1 - \left[1 - {\rm e}^{-p_k/ \sigma_{n}^{2}} \right]^{N^*}.
\end{equation}
Through this quantity it is possible to fix a detection threshold  $L_{{\rm Fa}}$,
\begin{equation} \label{eq:threshold}
L_{{\rm Fa}} = - \sigma_n^2 \ln\left[1 - (1 - \alpha)^{1/{N^*}} \right],
\end{equation}
corresponding to the level that one or more peaks due to the noise would exceed with a prefixed probability $\alpha$ when a number $N^*$ of ({\it statistically independent}) frequencies are inspected.  
Threshold $L_{{\rm Fa}}$ is called the {\it level of false alarm}.

For a periodic component with amplitude $A$, phase $\phi_l$ and frequency $f_l$ (in units of $1 / \Delta t$) in the set of the Fourier frequencies $\{ f_k \}$, $s_j = A \sin{(2 \pi f_l t_j + \phi_l)}$,
the periodogram will show a prominent peak  at $k=l$. Indeed, since $\xh_{N-k+1}$ is the complex conjugate of  $\xh_k$, then $\cos{[2 \pi (N-k+1) j]} = \cos{[2 \pi k j]}$ and  $\sin{[2 \pi (N-k+1) j]} = -\sin{[2 \pi k j]}$. Hence,
Eq.~(\ref{eq:dfti}) can be written in the form \citep{chu08}
\begin{equation} \label{eq:model}
x_j = \frac{1}{N} \sum_{k=0}^{N-1} a_k \cos{\frac{2 \pi k j}{N}} + b_k \sin{\frac{2 \pi k j}{N}},
\end{equation}
where
\begin{align}
a_k & = \sum_{j=0}^{N-1} x_j \cos{\frac{2 \pi k j}{N}}; \label{eq:ak} \\
b_k & = \sum_{j=0}^{N-1} x_j \sin{\frac{2 \pi k j}{N}}, \label{eq:bk}
\end{align} 
or
\begin{align}
a_k &= \frac{\xh_k + \xh_{N-k+1}}{2}; \label{eq:ak} \\
b_k & = i \frac{\xh_k - \xh_{N-k+1}}{2}. \label{eq:bk}
\end{align} 
Now, since
\begin{equation}
s_j = a_l \cos{\frac{2 \pi l j}{N}} + b_l \sin{\frac{2 \pi l j}{N}},
\end{equation}
only the coefficients $\xh_l$ and $\xh_{N-l}$ and hence only $\ph_l = (a_l^2 +b_l^2)/N$ will  be different from zero. More generally, if $x_j = A \sin{(2 \pi f^*_l t_j + \phi)} + n_j$, with
$f^*_l$ close but not identical to the Fourier frequency $f_l$, the periodogram takes the form of a squared ``$\rm{sinc}$'' function centered at $f_l^*$. Also in this case, it is expected that  $p_l >  L_{{\rm Fa}}$ for small values of 
$\alpha$ (typically $0.05$ or $0.01$). If $s(t)$ is semi-periodic or even non-periodic, the situation is more complicated since more peaks are expected, but the basic idea does not change.

Regular sampling has many advantages, among them: 
\begin{itemize}
\item The sine and cosine modes corresponding to the {\it Fourier frequencies} constitute an orthonormal basis for signal $\{ x_j \}$. This makes operations such as noise filtering, separation and/or detection of components of interest easier; 
\item The spectrogram can be shown to derive from the least-squares fit of model~(\ref{eq:model}) to the observed signal \citep[e.g. see][]{vio10}. This provides a physical interpretation of the quantity $p_k$ as energy associated with
the component at frequency $f_k$; 
\item Under the pure noise hypothesis $x_j = n_j$ and independently of $k$, $a_k $ and $b_k$ are uncorrelated (independent) Gaussian quantities.  As a consequence $p_k$ contains all the available information. In other words, 
the use of the joint distribution of $a_k$ and
$b_k$ does not provide any advantage with respect to the use of $p_k$. Moreover, the quantities $\{ p_k \}_{k=0}^{N/2}$ are mutually independent and have a known PDF. All of these facts permit the development  of simple and effective 
detection techniques; \\
\item Quite efficient algorithms are available for the computation of $\{ p_k \}$.
\end{itemize}
At the same time, however, it is necessary to stress that:
\begin{itemize}
\item The Fourier frequencies have no particular physical meaning. They constitute kinds of {\it natural frequencies} that, however, are intrinsic to the sampling characteristics and not to the signal under analysis.
This implies that the frequency of interest could not belong to such a set; 
\item If $x_j$ contains a sinusoidal component with frequency  $f_u > f_{{\rm Ny}} = 0.5$  (in units of $1/\Delta t$), the periodogram will show a peak in correspondence to a frequency $f = {\rm mod}{(f_u, 2 \pi)} < f_{{\rm Ny}}$ 
\footnote{The function $z={\rm mod}(x,y)$ provides the remainder $z$ from the division of $x$ by $y$.}. This puts an upper limit $f_{{\rm Ny}}$, the so called {\it Nyquist frequency},
on the maximal frequency that can be detected in a time series.
\end{itemize}
In conclusion, a regular sampling simplifies the analysis of the data as well as the development of efficient algorithms. However, especially in the context of exploratory data analysis, it suffers of some annoying limitations. 

\section{Periodogram analysis of irregularly sampled signals} \label{sec:irregular}

\subsection{Statistical issues} \label{sec:statistical}
In Astronomy often the experimental conditions do not permit a regular sampling of signals and this leads to the following.
First, it is no longer possible to define a set of 
{\it natural} frequencies (such as the Fourier frequencies) for which to compute the periodogram. Hence, there is no  reason for the number $N$ of frequencies to be equal to
the number $M$ of the sampling time instants $t_0, t_1, \ldots, t_{M-1}$. Therefore, we write the  transformation corresponding to that given 
by Eq.~(\ref{eq:dft}) in the general form
\begin{equation} \label{eq:dftirr}
\xh_f = \sum_{j=0}^{M-1} x_{t_j} {\rm e}^{-i 2 \pi f t_j },
\end{equation}
where, without loss of generality, we have $t_1=0$.
The spectrogram is still defined as $p_f = \left| \xh_f  \right|^2/M$. Similarly, Eqs.~(\ref{eq:ak})-(\ref{eq:bk}) become
\begin{align}
a_f & = \sum_{j=0}^{M-1} x_{t_j} \cos{2 \pi f t_j}; \label{eq:aki} \\
b_f & = \sum_{j=0}^{M-1} x_{t_j} \sin{2 \pi f t_j}, \label{eq:bki}
\end{align}
and 
\begin{equation} \label{eq:pf}
p_f = \frac{a_f^2 + b_f^2}{M}.
\end{equation}  

Second, the quantity $p_f$ loses its physical meaning and it no longer provides the energy of a signal at frequency $f$. Indeed, for a given  $f$, $p_f$ can be obtained
from the least-squares problem \citep{sto09}
\begin{align}
p_f & = \frac{1}{M} | \tilde{\beta}_f |^2, \label{eq:ls1} \\
\tilde{\beta}_f & =  \underset{ \beta_f}{\arg\min} \left[ \sum_{j=0}^{M-1} |x_{t_j} - \beta_f {\rm e}^{i 2 \pi f t_j}|^2 \right],  \label{eq:ls2} 
\end{align}
since it is readily verified that $\tilde{\beta}_f = \xh_f$.
If $\beta_f$ is expressed in the polar form $\beta_f = |\beta_f| {\rm e}^{i 2 \pi \phi_f}$, then
the least-squares problem~(\ref{eq:ls2}) can be rewritten in the form
 \begin{multline} \label{eq:ls3} 
\tilde{\beta}_f =  \underset{ \beta_f}{\arg\min} \left[ \sum_{j=0}^{M-1} [x_{t_j} - |\beta_f|  \cos{(2 \pi f t_j + \phi_f)}]^2 \right. + \\ \left. |\beta_f |^2 \sum_{j=0}^{M-1} \sin^2{(2 \pi f t_j + \phi_f)} \right].
\end{multline}
The first term in this equation represents the least-squares fit of a sinusoidal function, and it can have a physical meaning. The second term represents a data-independent quantity with no meaning in the context of the
model fit. Therefore, Eq.~(\ref{eq:ls3})  indicates that in the case of irregular sampling the periodogram is not equivalent to the least-squares fit of sinusoidal functions \citep[see also][]{vio10}. Consequently, the 
coefficients $a_f$ and $b_f$ given by Eqs.~(\ref{eq:aki})-(\ref{eq:bki}) do not provide the corresponding amplitudes. Since Eq.~(\ref{eq:dftirr}) can be interpreted as the correlation between $x_{t_j}$ and 
the sine and cosine modes with frequency $f$, {\it the periodogram  becomes a simple statistical measure of similarity between the experimental time series and a discrete sinusoidal signal of frequency $f$}.

Another issue linked to the irregular sampling is the fact that, even under the hypothesis of a noise signal with  $M=N$, although still with a Gaussian PDF,  $a_f$ and $b_f$ are no longer uncorrelated. As a consequence, 
the quantities $p_f/ \sigma^2_n$ no longer have an exponential PDF. This problem has been solved by \citet{lom76} and \citet{sca82}. Their approach, however, is a bit tortuous. 
A more intuitive, though equivalent, method is based on the least-squares model \citep{sto09, vio10}:
\begin{equation} \label{eq:lsmodel}
(\tilde{a}_f, \tilde{b}_f) =  \underset{ a_f, b_f}{\arg\min} \sum_{j=0}^{M-1} [x_{t_j} - a_f \cos{(2 \pi f t_j )} - b_f \sin{(2 \pi f t_j )}]^2.
\end{equation}
The solution of this problem is
\begin{equation}
\left(
\begin{array}{c}
\tilde{a}_f \\
\tilde{b}_f 
\end{array}
\right) = \Rb_f^{-1} \rb_f,
\end{equation}
where
\begin{align} 
\Rb_f & = \sum_{j=0}^{M-1}
\left(
\begin{array}{c}
\cos{(2 \pi f t_j)} \\
\sin{(2 \pi f t_j)} 
\end{array}
\right) 
\left(
\begin{array}{cc}
\cos{(2 \pi f t_j)} & \sin{(2 \pi f t_j)}
\end{array}  \right),  \\
\rb_f & = \sum_{j=0}^{M-1} \label{eq:rf}
\left(
\begin{array}{cc}
\cos{(2 \pi f t_j)} \\
\sin{(2 \pi f t_j)}
\end{array}
\right)  x_{t_j}.
\end{align}
The energy $\overline{p}_f$ associated with frequency $f$ is given by
\begin{align}
\overline{p}_f & = \sum_{j=0}^{M-1} \label{eq:power1}
\left(
\begin{array}{cc}
\left( \begin{array}{cc} \tilde{a}_f &   \tilde{b }_f  \end{array} \right)  & \left( \begin{array}{c} \cos{(2 \pi f t_j)} \\  \sin{(2 \pi f t_j)} \end{array} \right)
\end{array}
\right)^2,  \\
& = \left( \begin{array}{cc} \tilde{a}_f &   \tilde{b }_f  \end{array} \right) \Rb_f \left(\begin{array}{c} \tilde{a}_f \\ \tilde{b}_f  \end{array} \right),  \label{eq:power2} \\ 
& = \rb_f^T \Rb_f^{-1} \rb_f.   \label{eq:power3}
\end{align}
In the case of a time series of a Gaussian, zero-mean, white-noise $\{ n_{t_j} \}_{j=0}^{M-1}$ with variance $\sigma^2_n$, from Eq.~(\ref{eq:rf}) it is easily verifiable that the entries of the array $\rb_f$ are Gaussian, zero-mean, 
random quantities with covariance matrix 
$\sigma^2_n \Rb_f$.  Since $\Rb_f$ is a positive definite matrix, it can be factorized in the form $\Rb_f = \Rb_f^{1/2} \Rb_f^{1/2}$ with  $\Rb_f^{1/2}$ the {\it Cholesky factorization} of $\Rb_f$ \citep{bjo96}.
Therefore, the entries of the array $\rb_f^* = \Rb_f^{-1/2} \rb_f / \sigma^2_n$ are independent Gaussian random quantities with unit variance and the PDF  of
$\overline{p}_f / \sigma^2_n$ is the exponential distribution.
However, there is no guarantee that, whenever $f \ne f'$, $\overline{p}_{f}$ is independent of $\overline{p}_{f'}$. In general it is not, since with the least-squares model~(\ref{eq:lsmodel}) a single sinusoid of frequency $f$ is fitted
per time. As a consequence, in the expression for the threshold $L_{{\rm Fa}}$
as given by Eq.~(\ref{eq:threshold}), the number of frequencies $N$ should be substituted by the number $N_f \le N$  of independent frequencies. The point is that $N_f$ is not known in advance and in principle, $N_f$ can be
obtained from the rank of the covariance matrix  $\Rb_f$. This last procedure can be computationally quite expensive.
However, as stressed by \citet{sca82}, the dependence of $L_{{\rm Fa}}$ on $N_f$ is rather weak and in many situations, $N_f = M/2$ provides a reasonable choice \citep[e.g. see][]{vio10}.

Before concluding this section, a final remark concerns the advisability of working with mean-subtracted signals. If the mean value $\overline{x}$ of a 
signal is different from zero,  Eqs.~(\ref{eq:aki}), (\ref{eq:bki}) imply that its contributions $\overline{a}_f$ and $\overline{b}_f$ to the coefficients $a_f$ and $b_f$ are given by
\begin{align}
\overline{a}_f & = \overline{x} \sum_{j=0}^{M-1} \cos{2 \pi f t_j}; \\
\overline{b}_f & = \overline{x} \sum_{j=0}^{M-1} \sin{2 \pi f t_j}.
\end{align}
From these equations it appears that, independently of $f$, both $\overline{a}_f$ and $\overline{b}_f$ are different from zero. In other words, $\overline{x}$ influences the entire periodogram and not only in correspondence of the 
frequency $f=0$ as in the case of a regular sampling. Moreover, the contribution is different for distinct frequencies and, since for a given $f$ it is  ${\rm E}[\overline{a}_f \overline{b}_f] \ne 0$, with ${\rm E}[.]$ the
expectation operator, a spurious correlation is introduced 
between $a_f$ and $b_f$. Obviously, all that makes more complicated the spectral analysis of the signal of interest. Actually,  {\it if the time series are not too short}, the mean-subtraction operation does not imply particular problems.
In other cases case, modifications such as the ``floating-mean periodogram'' have to be used. For a detailed discussion of such a question see \citet{cum99, ree07, zec09, vio10}.

\subsection{Considerations about the Nyquist frequency }

Data with irregular sampling carry information that can be exploited in many ways. One of the benefits of an uneven sampling is the drastic reduction of the frequency aliasing (i.e. the aliasing of high frequencies down to 
lower ones). In other words, it is possible to identify periodic components with frequencies much higher than
the $f_{{\rm Ny}}$ corresponding to that of a time series with an identical number of equispaced data spanning the same time interval. When in Eq.~(\ref{eq:dft}) $k > N/2$, this frequency index can be written
as $k = N/2 + k'$. Then,
\begin{equation} \label{eq:ny}
\sin{\left( \frac{2 \pi k j}{N} \right)} \equiv \sin{\left( \pi j + \frac{2 \pi k' j}{N} \right)}= (-1)^j  \sin{\left(\frac{2 \pi k' j}{N} \right)}
\end{equation}
and similarly for the cosine function. Hence, $p_k = p_{k'}$. As a consequence, a sinusoidal component with frequency index  $k$ will produce a prominent peak in the periodogram also at $k' < k$. 
At the same time, a sinusoidal component with frequency index $k'$ will produce a prominent peak in the periodogram also at $k > k'$. Using a periodogram it is not possible to determine whether a sinusoidal
component is present in the signal with frequency index $k$ or $k'$. In the case of an irregular sampling Eq.~(\ref{eq:ny}) does not hold. This implies that periodogram can be used to distinguish a sinusoid with frequency $f$  from another 
one with frequency $f'$ also when $f' > f_{{\rm Ny}}$. In particular, \citet{eye99} found that, if the sampling time grid is in the form 
\begin{equation} \label{eq:sreg}
t_j = q_j \delta_t, 
\end{equation}
with $q_j$ integer numbers and $\delta_t$ the greatest common divisor for all $t_j$, then
\begin{equation} \label{eq:ny1}
f_{{\rm Ny}}=\frac{1}{2 \delta_t} \ge \frac{1}{2 \Delta t}. 
\end{equation}
This is explained as follows: if the sampling pattern is in the form given by Eq.~(\ref{eq:sreg}), then from $\{ x_{t_j} \}$ it is possible to obtain 
an even time series $\{ x'_l \}_{l=0}^{M M_q}$ where $M_q = t_{M-1}/\delta_t$ and
\begin{equation}
x'_l =
\begin{cases}
x_{t_j} & \text{if } l \delta_t = t_j,\\
0 & \text{otherwise}.
\end{cases}
\end{equation}
The Nyquist frequency for this time series is given by Eq.~(\ref{eq:ny1}). A formula for its calculation is given in \citet{koe06}.

\begin{figure*}
        \resizebox{\hsize}{!}{\includegraphics{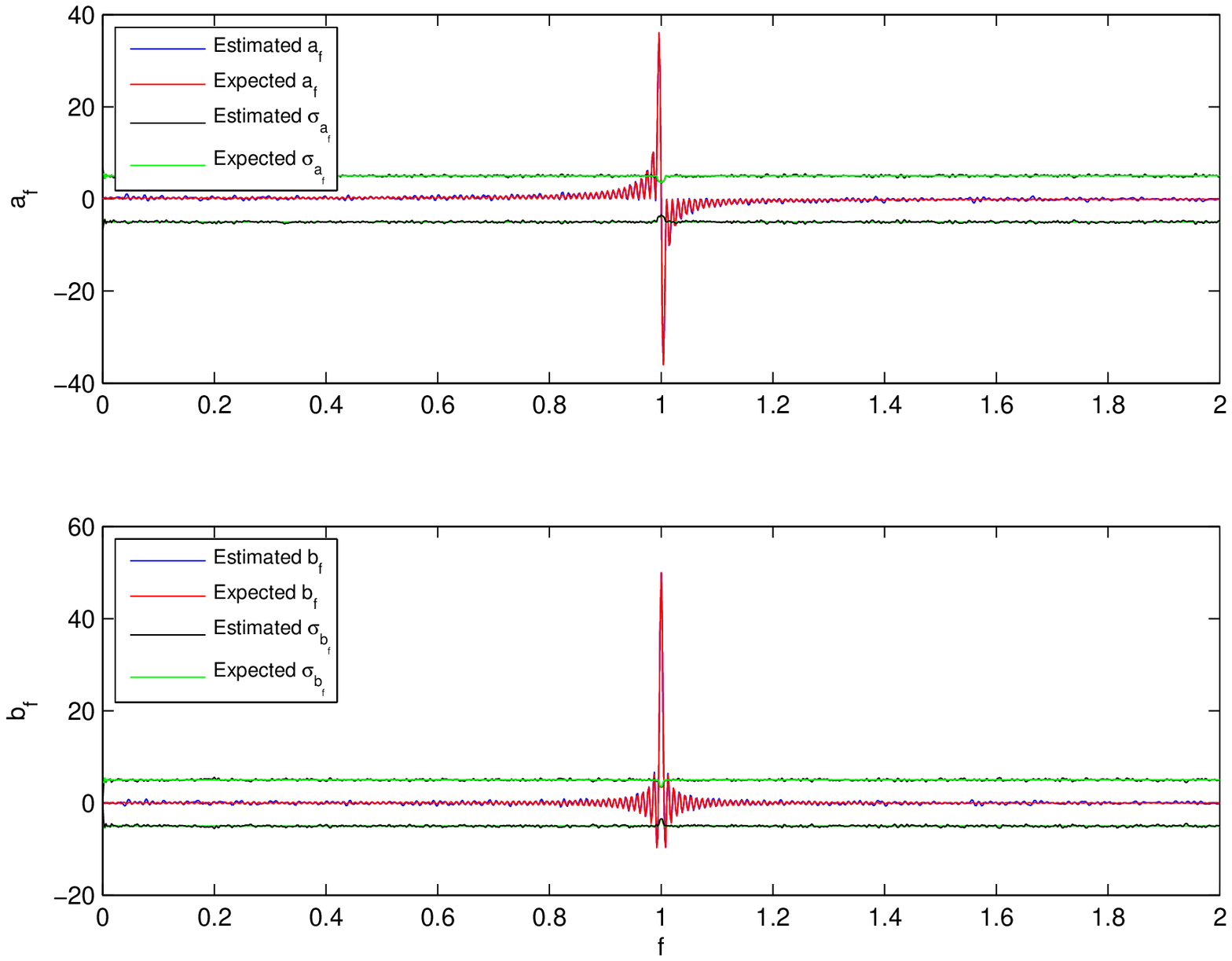}}
        \caption{Theoretical vs. estimated values of coefficients $a_f$,  $b_f$ (cf. Eqs.(\ref{eq:af}),  (\ref{eq:bf}), (\ref{eq:af0}), (\ref{eq:bf0}), and the corresponding standard deviations $\sigma_{a_f}$, $\sigma_{b_f}$), 
for a signal $x_{t_j}=\sin{(2 \pi f_0 t_j)}$, $j=0,1, \ldots, 99$, when the sampling 
time instants $t_j$ are uniformly and independently distributed in the interval $[0, 99]$. Here, $f_0 = 1$ in units of the mean $\Delta t$ ($=1$). The estimated quantities are based {\it on the mean of $500$ different numerical experiments}.}
        \label{fig:detection1}
\end{figure*}

\begin{figure*}
        \resizebox{\hsize}{!}{\includegraphics{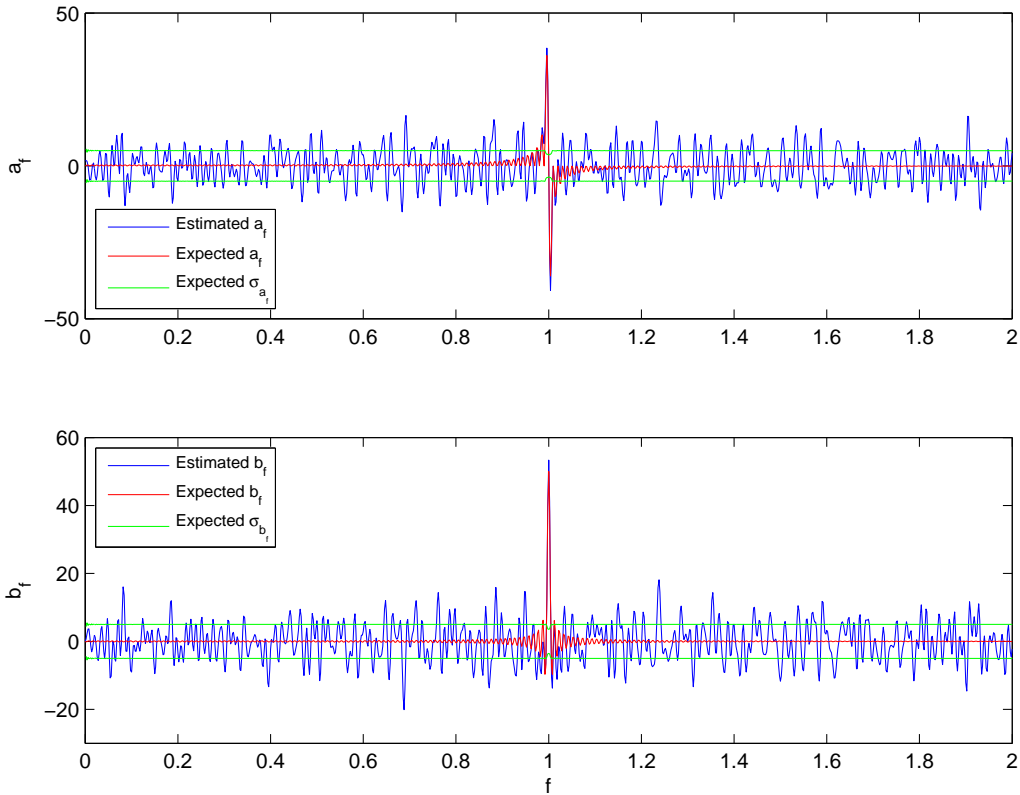}}
        \caption{Theoretical vs. estimated values of coefficients $a_f$, $b_f$ (cf. Eqs.~(\ref{eq:af}),  (\ref{eq:bf}), (\ref{eq:af0}), (\ref{eq:bf0}), and the corresponding standard deviations $\sigma_{a_f}$, $\sigma_{b_f}$), 
for a signal $x_{t_j}=\sin{(2 \pi f_0 t_j)}$, $j=0,1, \ldots, 99$, when the sampling 
time instants $t_j$ are uniformly and independently distributed in the interval $[0, 99]$. Here, $f_0 = 1$ in units of the mean $\Delta t$ ($=1$) and the  estimated quantities are based only {\it on a single numerical simulation}.}
        \label{fig:detection2}
\end{figure*}

\begin{figure*}
        \resizebox{\hsize}{!}{\includegraphics{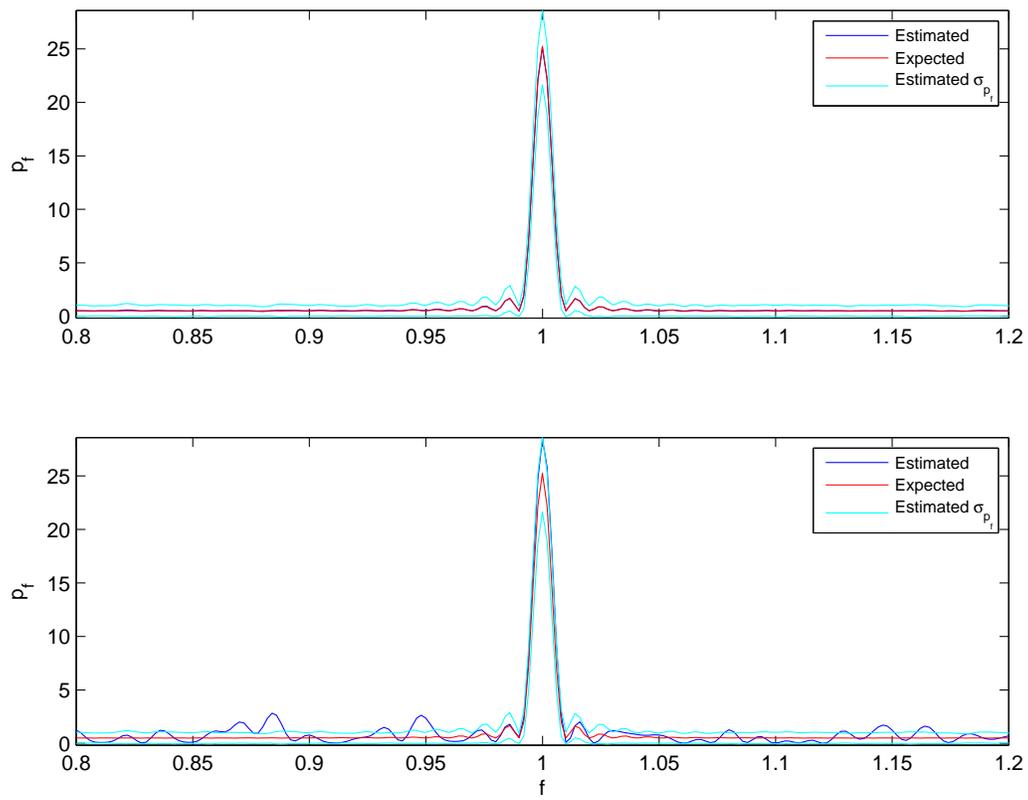}}
        \caption{Periodograms corresponding to Fig.~\ref{fig:detection1} (top panel) and Fig.~\ref{fig:detection2} (bottom panel).}
        \label{fig:power}
\end{figure*}

If the sampling pattern cannot be expressed in the form~(\ref{eq:sreg}), then $\delta_t = 0$. In this case there is the
surprising result that $f_{{\rm Ny}} = \infty$. For example, this happens when the sampling times are randomly and uniformly distributed in the interval $[0, T]$. If $x(t) = \sin{(2 \pi f_0 t + \phi)}$ it is possible to show that
the expected values of $a_f$ and $b_f$ are, respectively,
\begin{multline} \label{eq:af}
{\rm E}_t[a_f] = \frac{M}{2 T} \left\{ \frac{\cos{[2 \pi (f-f_0) T - \phi] - \cos{[\phi]}}}{2 \pi (f - f_0)} \right. - \\
\left. \frac{\cos{[2 \pi (f+f_0) T - \phi]}-\cos{[\phi]}}{2 \pi (f + f_0)} \right\},
\end{multline}
\begin{multline}  \label{eq:bf}
{\rm E}_t[b_f] = \frac{M}{2 T} \left\{ \frac{\sin{[2 \pi (f-f_0) T - \phi] + \sin{[\phi]}}}{2 \pi (f - f_0)} \right. - \\
\left. \frac{\sin{[2 \pi (f+f_0) T - \phi]}-\sin{[\phi]}}{2 \pi (f + f_0)} \right\},
\end{multline}
if $f \neq f_0$ and
\begin{equation}  \label{eq:af0}
{\rm E}_t[a_{f_0}] = \frac{M}{2 T} \left\{\frac{\cos{[\phi]} - \cos{[4 \pi f_0 T + \phi]}}{4 \pi f_0}  - T \sin{[\phi]}\right\},
\end{equation}
\begin{equation}  \label{eq:bf0}
{\rm E}_t[b_{f_0}] = \frac{M}{2 T} \left\{\frac{\sin{[\phi]} - \sin{[4 \pi f_0 T + \phi]}}{4 \pi f_0} + T \cos{[\phi]}\right\},
\end{equation}
if $f = f_0$.  For increasing values of T, ${\rm E}_t[a_f] \to - M \sin{(\phi)}/2$ and ${\rm E}_t[b_f] \to M \cos{(\phi)}/2$. The equations that provide the expected standard deviations 
$\sigma_{a_f}$ and $\sigma_{b_f}$ are horribly long but, for $T$ sufficiently large with respect to $f$, both these quantities are approximately equal to $\sqrt{M}/ 2$.
This implies that the uniform random sampling introduces a noise that, however, becomes rapidly 
negligible for increasing values of $M$. The remarkable point is that these results are independent of the frequency $f_0$. Hence, $f_0$ can be arbitrarily large. As an example, 
Fig.~\ref{fig:detection1} shows ${\rm E}_t[a_f]$  and ${\rm E}_t[b_f]$ together with the corresponding standard deviations $\sigma_{a_f}$ and $\sigma_{b_f}$  for the case where $f_0 = 1$  (i.e. twice the Nyquist frequency
corresponding to the mean sampling time step), $\phi=0$, 
$M=50$ and $T=M-1$. For comparison,  the results obtained from $500$ numerical simulations are also displayed.
In Fig.~\ref{fig:detection2} the theoretical ${\rm E}_t[a_f]$  and ${\rm E}_t[b_f]$ are compared with the result from a single simulation. Finally, in Fig.~\ref{fig:power} the corresponding periodograms are shown as well the
corresponding standard deviation as obtained from the numerical simulations (the expected values of this quantity give rise to terrible long equations).
From these results, one could argue that it is possible to detect a periodic component independently of its frequency. However, from the analysis of 
Eqs.~(\ref{eq:af})-(\ref{eq:bf}) it can be inferred that  the width of the peak at frequency $f_0$ is inversely proportional to $T$. As a consequence, if $T$ is large, the peak will be quite narrow and
there is a concrete risk of missing it if the periodogram is not computed for a sufficiently large number of frequencies.  In addition for very high frequencies, a periodogram can be deeply altered by even small  errors in 
the sampling times $t_j$ (see below).

\subsection{Computational issues} \label{sec:computation}

A difficulty introduced by an irregular sampling is the lack of efficiency of the algorithm for the computation of $\{ \xh_f \}$. Indeed, algorithms based on the fast Fourier transform (FFT) are inapplicable and the direct implementation 
of Eq.~(\ref{eq:dftirr}) requires an operation count of order $M N$ that is computationally quite inefficient. The solutions proposed for overcoming this problem are based
on algorithms/techniques that are not trivial \citep[e.g. ][]{pre07, kei08}, that make it difficult to deal with the experimental signals if the computation of  $\{ p_f \}$ or of the coefficients $\{ \xh_f \}$ represents only one step in the analysis procedure. 
For example, after filtering in the frequency domain, it could be necessary to Fourier invert the sequence $\{ \xh_f \}$.  In the case of irregular sampling, an inversion similar to that given by Eq.~(\ref{eq:dfti}) does not exist.
A simple solution that makes things easier consists of rebinning the original sampled signal onto an arbitrarily dense regular time grid. According to this approach, the time interval $[t_0, t_{M-1}]$ is divided into a number $\Mmatc - 1 \gg M$
of subintervals (bins) centered at $\{ \tau_l \}_{l=0}^{\Mmatc-1}$ time instants. A new time series $\chi_{\tau_0}, \chi_{\tau_1}, \ldots, \chi_{\tau_{\Mmatc-1}}$ is obtained by assigning each $t_j$ to the nearest bin, i.e. 
by setting  $\chi_{\tau_{l_j}} = x_{t_j}$ if $\tau_{l_j}$ is the time instant closest to $t_j$, and zero otherwise. More specifically, if an array $\{ \chi_{\tau_l} \}$ of $\Mmatc$ zeros is created,  index $l_j$ is given by 
\begin{equation}
l_j={\rm round} \left[ (\Mmatc - 1) \frac{t_j - t_0}{t_{M-1}-t_0} \right], 
\end{equation}
where ${\rm round}[t]$ is the operator that provides the integer closest to $t$.
In this way a grid of $\Mmatc$ time instants, regularly spaced with a time step $\Delta \tau = (t_{M-1}-t_0) /  (\Mmatc-1)$, is obtained but in the resulting time series $\{ \chi_{\tau_l} \}$ some of the entries are equal to zero. 
The FFT algorithm can be directly applied to this time series and the LS periodogram computed through Eqs.~(\ref{eq:ak})-(\ref{eq:bk}) and (\ref{eq:power1})-(\ref{eq:power3}).
Intuitively, this approach may be expected to provide satisfactory results if the differences $\{ \delta \tau_l \} = \{t_j - \tau_l \}$ are reasonably small with respect to the frequencies of interest.

To quantify this assertion, let us suppose, without loss of generality, that the signal under study is a sinusoid $x_{t_j} = \sin{(2 \pi f t_j+\phi)}$ 
which is rebinned in such a way as to obtain a time series $\chi_{\tau_l} = \sin{[2 \pi f (\tau_l +\delta \tau_l)+\phi]}$. Let suppose also that $\{ \delta \tau_l \}$ are randomly distributed in the interval  $[-0.5 \Delta \tau, 0.5 \Delta \tau]$ with ${\rm E}[\delta \tau_l]=0$.
From Eqs.~(\ref{eq:aki})-(\ref{eq:bki})
\begin{align}
a_f & = \sum_{l=0}^{\Mmatc - 1}  \sin{[2 \pi f (\tau_l +\delta \tau_l)+\phi]} \cos{(2 \pi f \tau_l)}; \label{eq:akr} \\
b_f & = \sum_{l=0}^{\Mmatc - 1}  \sin{[2 \pi f (\tau_l +\delta \tau_l)+\phi]} \sin{(2 \pi f  \tau_l)}. \label{eq:bkr}
\end{align} 
Now, if the terms $\sin{[2 \pi f (\tau_l +\delta \tau_l)+\phi]}$ are expanded up to the linear term, one obtains
\begin{align}
\tilde{a}_f &= \sum_{l=0}^{\Mmatc - 1}  [ \sin{(2 \pi f \tau_l+\phi)} + 2 \pi f \delta \tau_l \cos{(2 \pi f \tau_l+\phi)} ] \cos{(2 \pi f \tau_l)}; \\
\tilde{b}_f &= \sum_{l=0}^{\Mmatc - 1}  [ \sin{(2 \pi f \tau_l+\phi)} + 2 \pi f \delta \tau_l \cos{(2 \pi f \tau_l+\phi)}] \sin{(2 \pi f \tau_l)},
\end{align}
or
\begin{align} 
\tilde{a}_f &= a_f + \sum_{l=0}^{\Mmatc - 1}  2 \pi f \delta \tau_l \cos{(2 \pi f \tau_l+\phi)} \cos{(2 \pi f \tau_l)}; \label{eq:approx1} \\
\tilde{b}_f &= b_f + \sum_{l=0}^{\Mmatc - 1}  2 \pi f \delta \tau_l \cos{(2 \pi f \tau_l+\phi)} \sin{(2 \pi f \tau_l)}. \label{eq:approx2}
\end{align}
If the time grid  $\{ \tau_l \}$ is fixed it results that ${\rm E}_{\delta \tau}[\tilde{a}_f]=a_f$ and  ${\rm E}_{\delta \tau}[\tilde{b}_f]=b_f$. Moreover, if it is assumed that the quantities $\delta \tau_l$ 
are  distributed independently and identically from a uniform PDF as well as independent of $\{ \tau_l \}$, it happens that
\begin{align}
\sigma_{\tilde{a}_f} &=\frac{2 \pi f \Delta \tau}{\sqrt{12}} \sqrt{\sum_{l=0}^{\Mmatc - 1}  \cos^2{(2 \pi f \tau_l+\phi)} \cos^2{(2 \pi f \tau_l)}}; \label{eq:approx3} \\
\sigma_{\tilde{b}_f} &=\frac{2 \pi f \Delta \tau}{\sqrt{12}}  \sqrt{\sum_{l=0}^{\Mmatc - 1}  \cos^2{(2 \pi f \tau_l+\phi)} \sin^2{(2 \pi f \tau_l)}}. \label{eq:approx4}
\end{align}
As expected, from this result it is evident that the error introduced by the rebinning operation is proportional to the product of the frequency $f$ and the sampling time step $\Delta \tau$ . 
Hence, an accuracy to any desired precision can be obtained if $\Delta \tau$ is chosen sufficiently small. In practical applications, such choice does not represent a critical step: once the largest 
frequency $f_{\rm max}$ of interest (in units of $1 /\Delta \tau$) is set, it is sufficient that $\Delta \tau \ll 1/f_{\rm max}$.

\begin{figure*}
        \resizebox{\hsize}{!}{\includegraphics{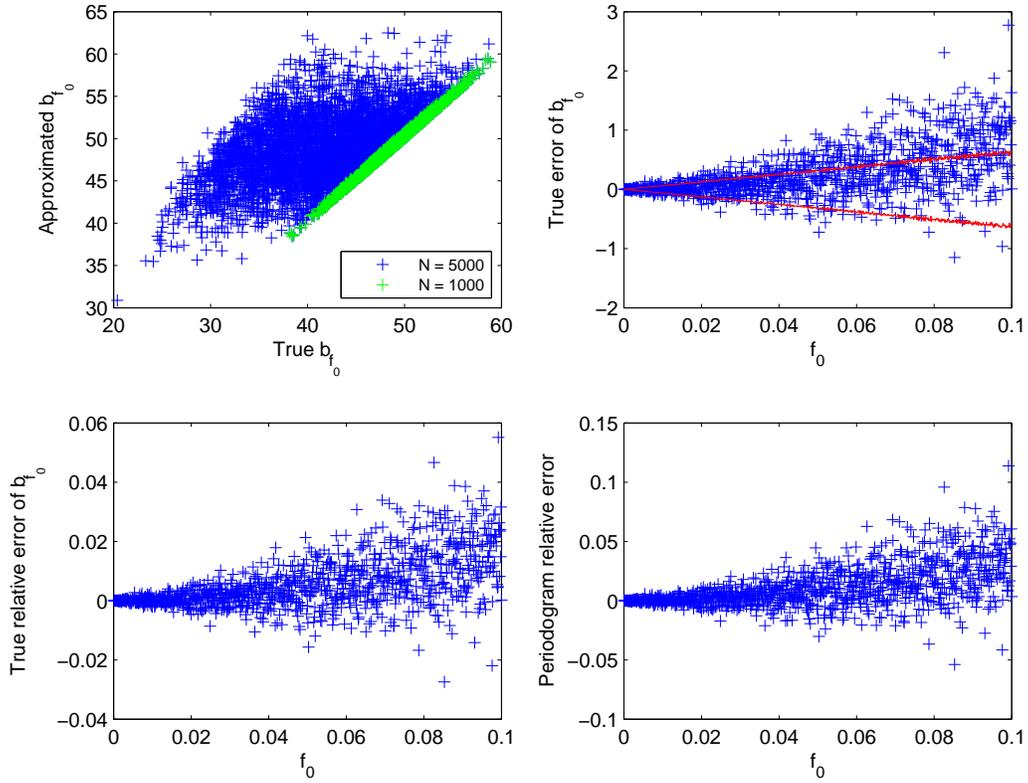}}
        \caption{Numerical experiment to test the effects of rebinning of an irregular time series on a regular time grid. Here,  $x_{t_j}=\sin{(2 \pi f_0 t_j)}$, $j=0,1, \ldots, 99$, with the sampling time instants independently
and uniformly distributed in the interval $[0, 10000]$ and then rounded to the nearest integer. A number $N=5 \times 10^3$ of equispaced frequencies are considered in the set $[1/N, 2/N, \ldots 0.5]$. 
Top-left panel: linearly approximated vs. true $b_{f_0}$. The first $10^3$ frequencies are plotted in green; Top-right panel: corresponding absolute errors. The expected standard deviation interval derived from the
linear approximation is in plotted in red; Bottom-left panel:
corresponding relative errors; Bottom-right panel: relative errors of the corresponding periodogram.}
        \label{fig:deltat}
\end{figure*}

For illustrative purposes, Fig.~\ref{fig:deltat} shows the results of a numerical simulation where a sinusoid $x_{t_j} = \sin{(2 \pi f_0 t_j)}$ is sampled on $100$ times $\{ t_j \}_{j=0}^{99}$ that are randomly and independently 
generated from a uniform distribution in the interval $T=[0, 10000]$ (in free units). The time instants $\{ t_j \}$ have been rebinned on a regular time grid $[0, 10000]$ by setting $\tau_l = {\rm round}[t_j]$.
Proceeding in this way, the time instant $\tau_l$ approximates the corresponding  $t_j$ with a precision of four digits and $\Delta \tau = 10^{-4}$. 
The frequency $f_0$ is considered in the interval $[10^{-4}, 0.1]$ in units of $(\Delta \tau)^{-1}$. 
From this figure it is evident that the linear approximation in Eqs.~(\ref{eq:approx1})-(\ref{eq:approx4}) holds up to frequencies of about $0.01$. However, both the approximated coefficients $\{ b_{f_0} \}$ as well the approximated periodogram
$p_{f_0}$ are within some percent with respect to the true value up to a frequency of $0.1$  (for the coefficients $\{ a_f \}$ similar results hold). It is worth stressing that $f_0 = 0.1$ is a rather high frequency with respect to a mean $\Delta t \approx 100$.

\section{Is the Lomb-Scargle periodogram really advantageous?}  \label{sec:spectrogram}

In Sec.~\ref{sec:computation} it has been shown that the LS periodogram can be computed with accuracy to any desired precision without the necessity of dedicated algorithms/software. At this point, assuming that the error of the approximation 
is negligible or even that an exact algorithm has been used, one can go one step further and wonder whether, to test
the statistical significance of a peak, the decorrelation of the coefficients $a_f$ and $b_f$, which is at the heart of the LS periodogram, is really a necessary operation. Using arguments based on the {\it spectral windows}
$W_f = \sum_{j=0}^{M-1} \exp{(-2 \pi f t_j)}$, \citet{vio10} have suggested that this is not the case since the correlation coefficient $\rho$ between $a_f$ and $b_f$ 
is typically close to zero. Here, to support this claim we follow a different approach. If $x_{t_j} \equiv \{ n_{t_j} \}$, with $\{ n_{t_j} \}$
the realization of a discrete, zero-mean white-noise process with standard deviation $\sigma_n$, then ${\rm E}_n[a_f]=0$, ${\rm E}_n[b_f]=0$ and
\begin{equation} \label{eq:rho}
\rho = \frac{{\rm E}_n[a_f b_f]}{ \sigma_n^2 \sqrt{\sum_{j=0}^{M-1} \cos^2{(2 \pi f t_j)}}  \sqrt{\sum_{j=0}^{M-1} \sin^2{(2 \pi f t_j)}}},
\end{equation}
where
\begin{align}
{\rm E}_n[a_f b_f] & = \sigma_n^2 \sum_{j=0}^{M-1} \cos{(2 \pi f t_j)} \sin{(2 \pi f t_j)}, \label{eq:corr1} \\ 
                          & = \frac{\sigma_n^2}{2} \sum_{j=0}^{M-1} \sin{(4 \pi f t_j)}. \label{eq:corr2}
\end{align} 
In the regular sampling case, it results that ${\rm E}_n[a_f b_f]=0$  and consequently $\rho = 0$. The same does not hold in the irregular sampling case.  However,
since $ \sin{(4 \pi f t_j) }$ is an odd function, one may expect that ${\rm E}_n[a_f b_f] \approx 0$ and hence $\rho \approx 0$  if the angles $\{ \alpha_j \} = 4 \pi f \{ t_j \}$ of a unit circle are uniformly and/or symmetrically distributed. In practical applications 
this condition 
is not infrequently met. For example, in the case of $M$ independent sampling time instants randomly and uniformly distributed in the interval $[0, T]$, one finds that the expected correlation coefficient $\rho_{t, a_f b_f}$ is given by
\begin{align}
\rho_{t, a_f b_f} &= \frac{{\rm E}_t\{{\rm E}_n[a_f b_f]\}}{\sigma_{t, a_f} \sigma_{t, b_f}}; \\
&= \frac{1 - \cos{(4 \pi f T)}}{\sqrt{(4 \pi  f T)^2 - \sin^2{(4 \pi f T)}}},
\end{align}
where $\sigma_{t, a_f}$ and $\sigma_{t, b_f}$ denote the standard deviations with respect to the time instants $\{ t_j \}$.
From this equation it is clear that $\rho_{t, a_f b_f}$ goes rapidly to zero for increasing values of $T$. For fixed times, a formal proof is difficult since it is strictly dependent on the specific sampling pattern. 
However, it is  improbable that the combination of the frequencies $f$ and the times $\{ t_j \}$ makes the distribution of the angles $\{ \alpha_j \}$ strongly nonuniform and/or asymmetric. For example,
high values of $\rho$ can be obtained if all the angles $\alpha_j$ are distributed in an interval $[\alpha^*-\epsilon  \pi, \alpha^*+\epsilon \pi] \subseteq [0, \pi] $ with $ \alpha^* \in [0, \pi]$ a given angle and
$\epsilon$ a real number that takes its value in the interval $\min{[\alpha^*/\pi, 1-\alpha^*/\pi]}$. The condition for this to happen is
\begin{equation}
4 \pi f t_j = \alpha^* + [2 \kappa \pi \pm \epsilon \pi],
\end{equation}
with $\kappa$ an integer, or
\begin{equation} \label{eq:irrs}
t_j = \frac{\alpha^*}{4 \pi f} + \frac{1}{2f}\left[\kappa \pm \frac{\epsilon}{2} \right].
\end{equation}
From this equation it results that 1) the sampling pattern must be constituted by times distributed in equispaced time intervals with the same duration which is proportional to $\epsilon$; 2) such a sampling pattern is specific to each
frequency $f$. This means that, even in the case where the sampling is such as to produce a high $\rho$ for a given frequency, the same could not be true for other frequencies. This is not a rigorous demonstration
of the fact that a nonuniform and irregular distribution of the angle $\alpha_j$ is improbable. In fact, combinations of sampling times, lags and frequencies are possible that can do the job. However, the considerations above
suggest that things have to conspire to produce remarkable effects. 

\begin{figure*}
        \resizebox{\hsize}{!}{\includegraphics{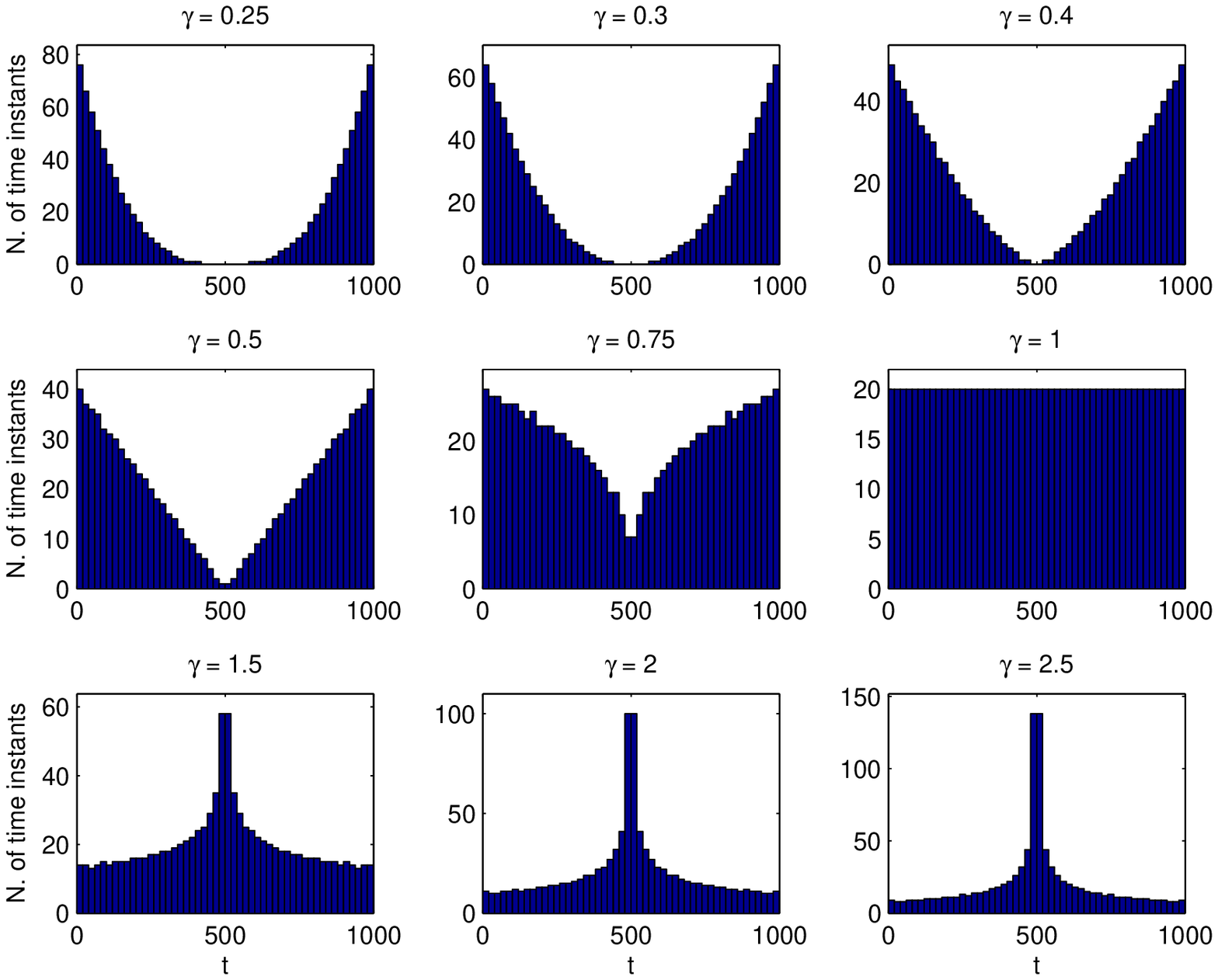}}
        \caption{Distribution of the first set of irregular sampling time instants used to test the effects of the rebinning operation on the accuracy of the computed Lomb-Scargle periodogram. 
	When $\gamma=1$ the sampling is regular and becomes more and more irregular when $\gamma  \to 0$ or $\gamma \to \infty$. Here the case with $M=1000$ sampling time instants is shown. The distribution for the case $M=100$ is similar.}
        \label{fig:time1}
\end{figure*}

\begin{figure*}
        \resizebox{\hsize}{!}{\includegraphics{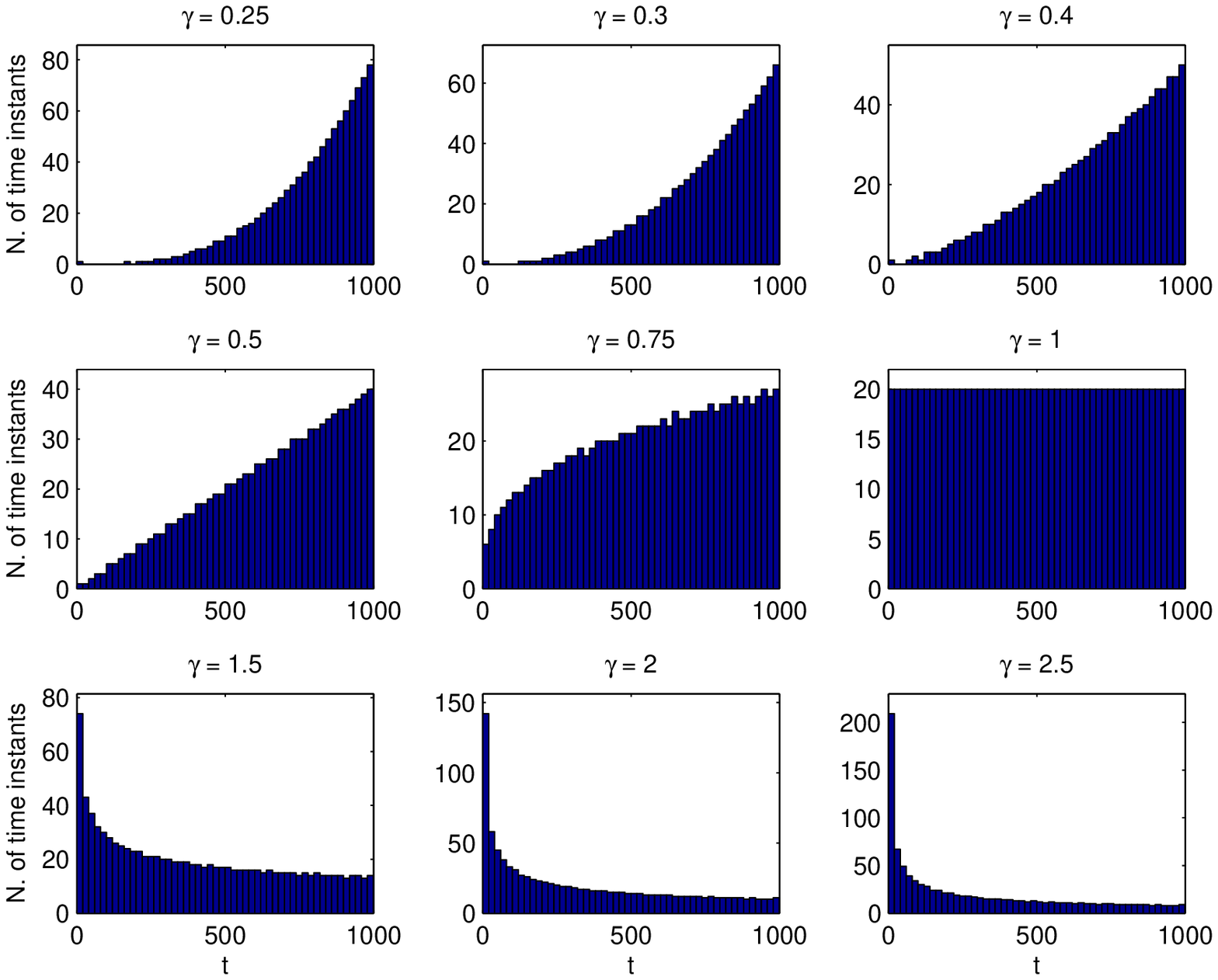}}
        \caption{Distribution of the second set of irregular sampling time instants used to test the effects of the rebinning operation  on the accuracy of the computed Lomb-Scargle periodogram.
	When  $\gamma=1$ the sampling is regular and becomes more and more irregular when $\gamma  \to 0$ or $\gamma \to \infty$. Here the case with $M=1000$ sampling time instants is shown. The distribution for the case $M=100$ is similar.}
        \label{fig:time2}
\end{figure*}

\begin{figure*}
        \resizebox{\hsize}{!}{\includegraphics{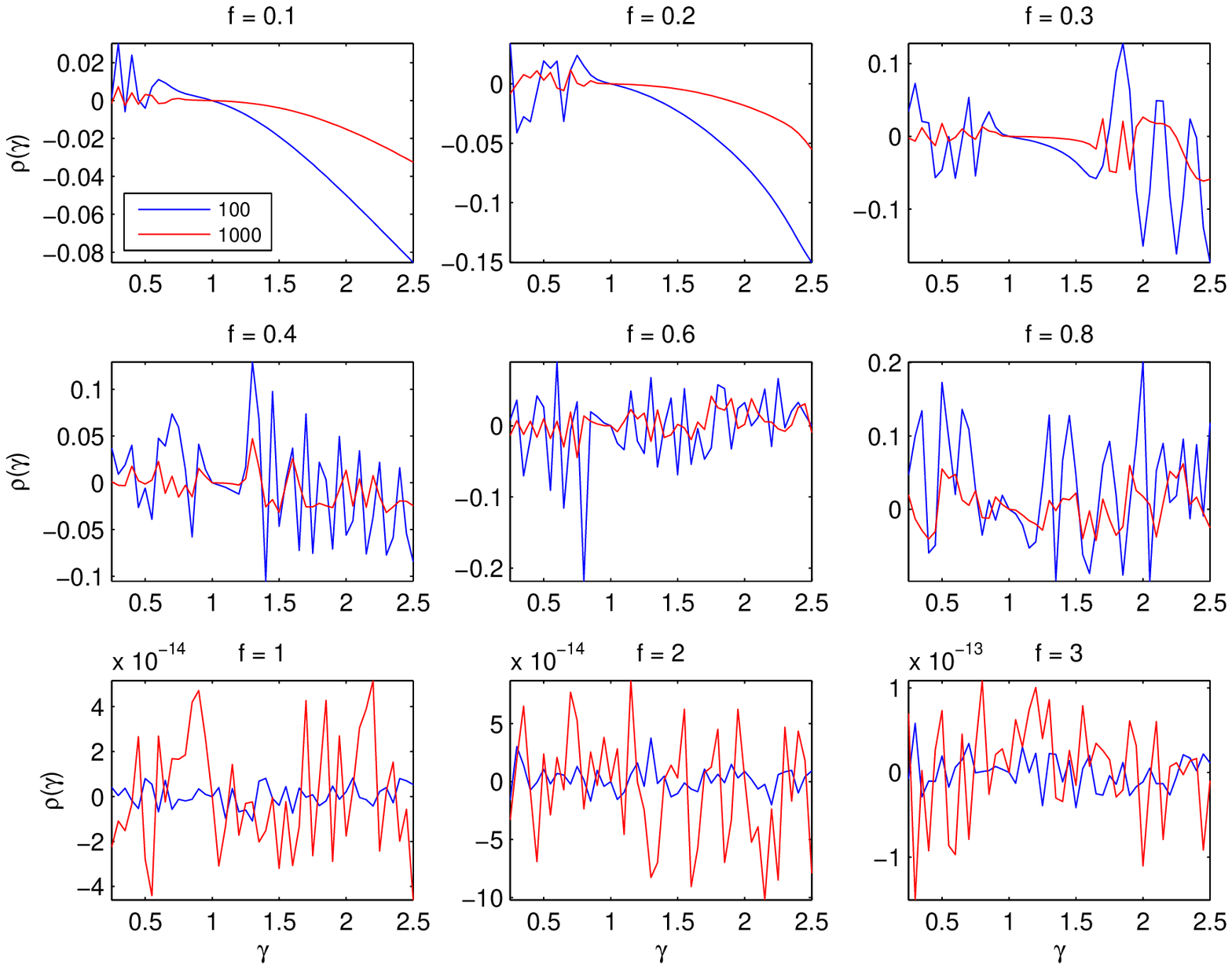}}
        \caption{Correlation coefficients $\rho$, cf. Eq.~(\ref{eq:rho}), of $a_f$ with  $b_f$ (cf. Eqs.(\ref{eq:akr}), (\ref{eq:bkr})), against the $\gamma$ parameter for a set of different frequencies $f$ and a number of sampling time instants $M=100$ 
        (blue line) and $M=1000$ (red line), distributed as shown in Fig.~\ref{fig:time1}. In spite of the extremely irregular sampling, significant $\rho$ occurs only for small data sets. But even in this case the correlation is weak.}
        \label{fig:corr1}
\end{figure*}

\begin{figure*}
        \resizebox{\hsize}{!}{\includegraphics{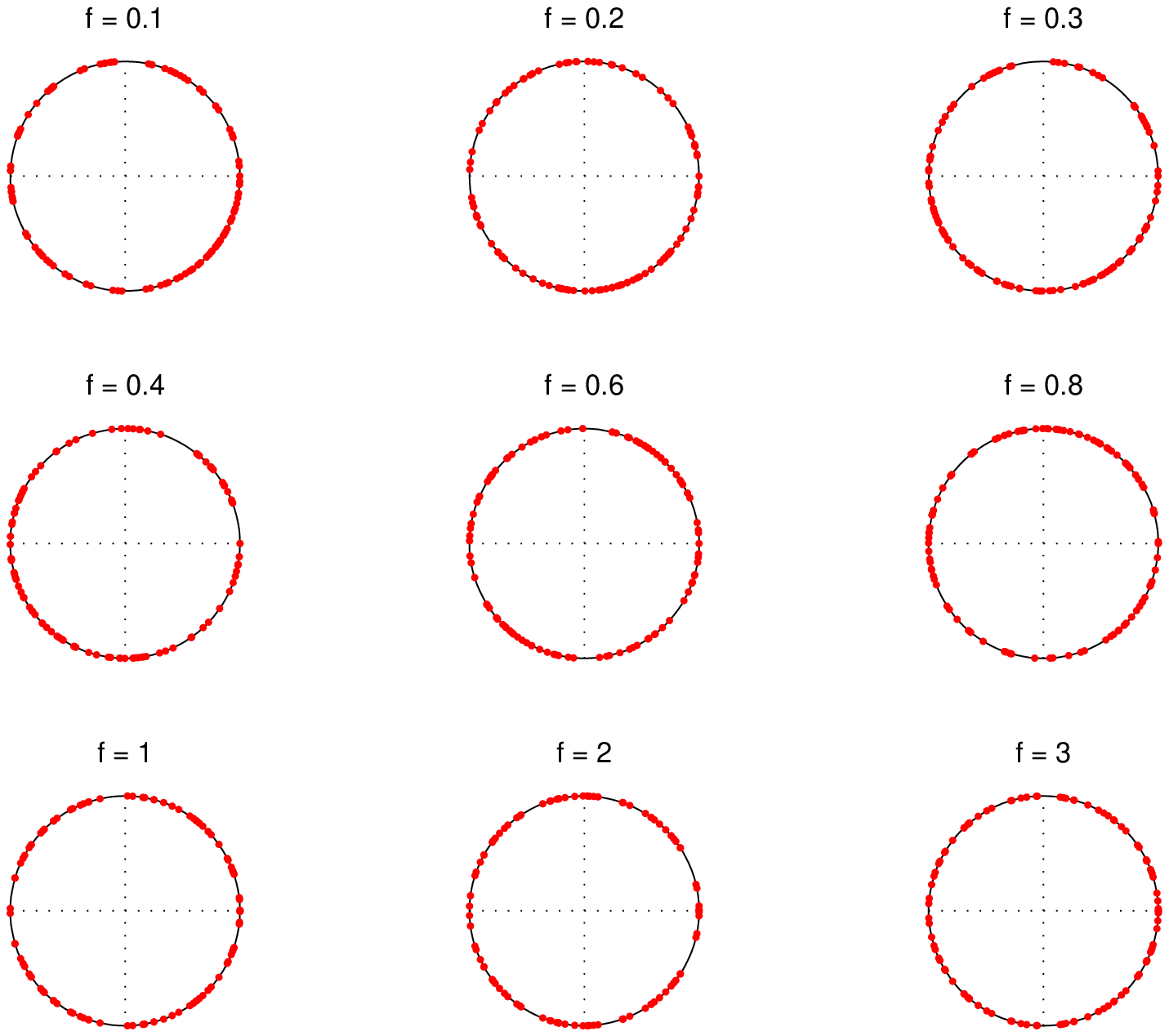}}
        \caption{Distribution of the angles $\alpha_j$ on the unit sphere for the set of  frequencies $f$ as in Fig.~\ref{fig:corr1} and a number of sampling time instants $M=100$ distributed as in the bottom-right panel of Fig.~\ref{fig:time1}.}
        \label{fig:angles1}
\end{figure*}

\begin{figure*}
        \resizebox{\hsize}{!}{\includegraphics{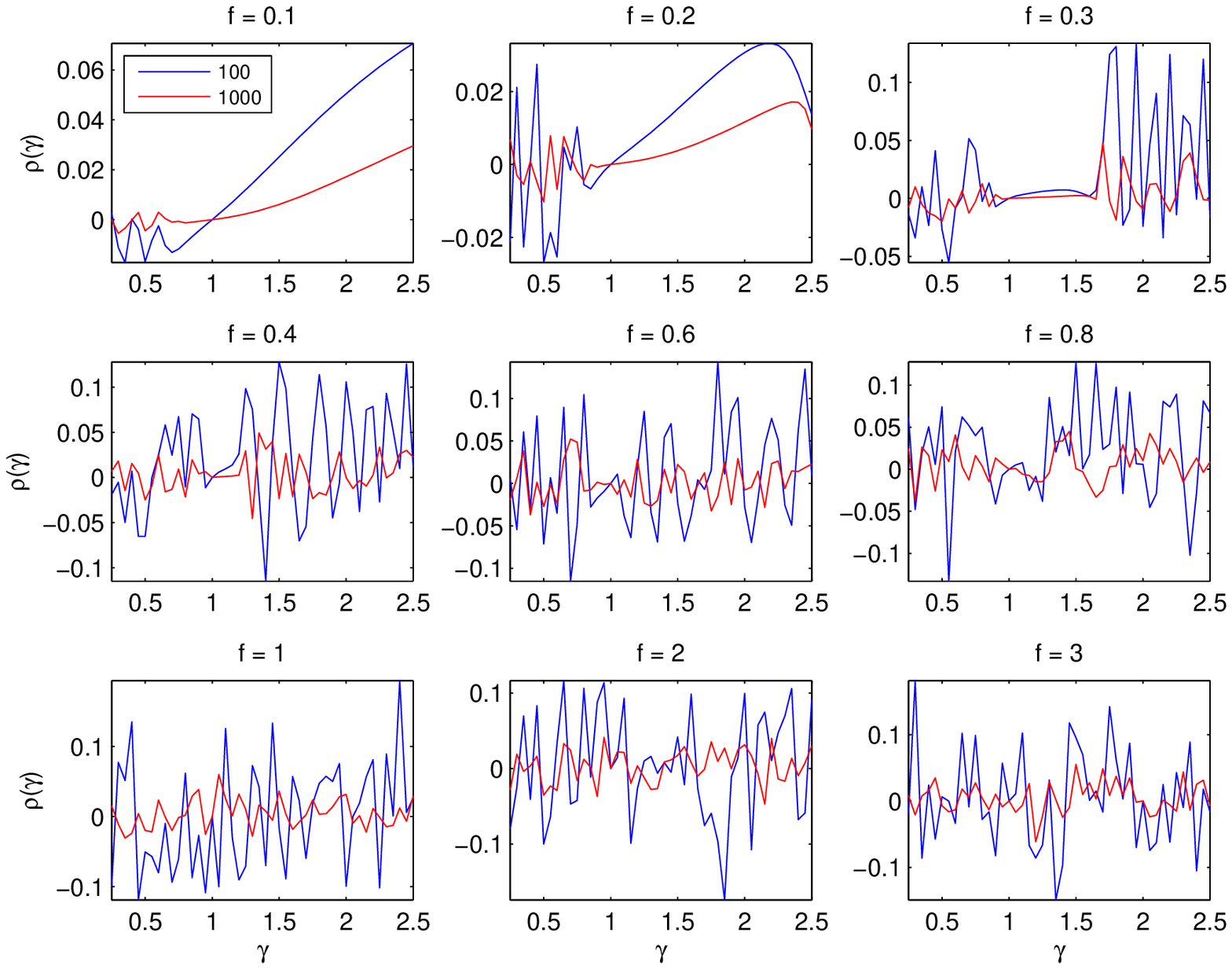}}
        \caption{Correlation coefficients $\rho$ (cf. Eq.~(\ref{eq:rho})), of $a_f$ with $b_f$ (cf. Eqs.(\ref{eq:akr}), (\ref{eq:bkr})), against the $\gamma$ parameter for a set of different frequencies $f$ and a number of sampling time instants $M=100$ 
        (blue line) and $M=1000$ (red line), distributed as shown in Fig.~\ref{fig:time2}. In spite of the extremely irregular sampling, significant $\rho$ occurs only for small data sets. But even in this case the correlation is weak.}
        \label{fig:corr2}
\end{figure*}

\begin{figure*}
        \resizebox{\hsize}{!}{\includegraphics{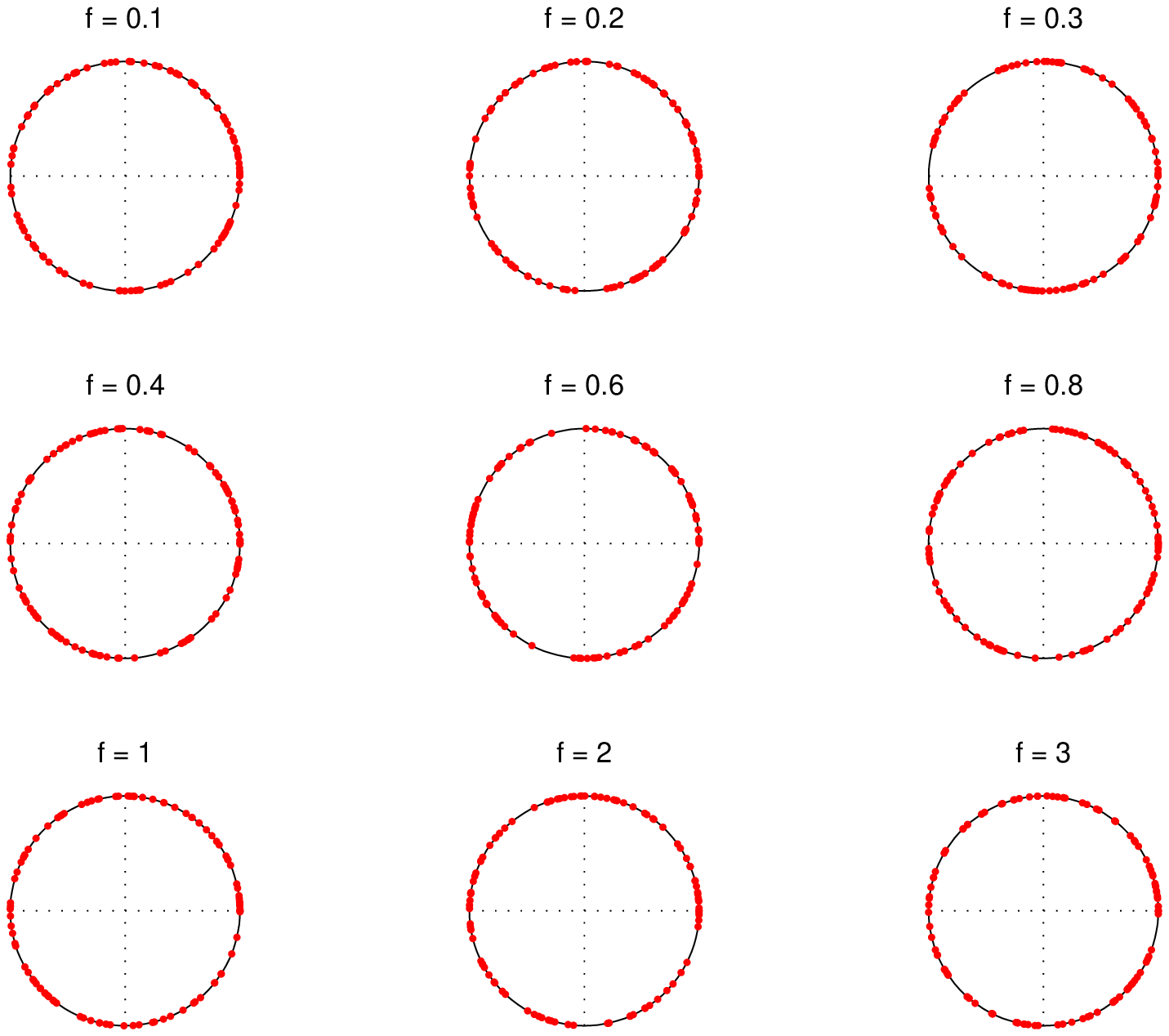}}
        \caption{Distribution of the angles $\alpha_j$ on the unit sphere for the set of  frequencies $f$ as in Fig.~\ref{fig:corr2} and a number of sampling time instants $M=100$ distributed as in the bottom-right panel of Fig.~\ref{fig:time2}.}
        \label{fig:angles2}
\end{figure*}

\begin{figure*}
        \resizebox{\hsize}{!}{\includegraphics{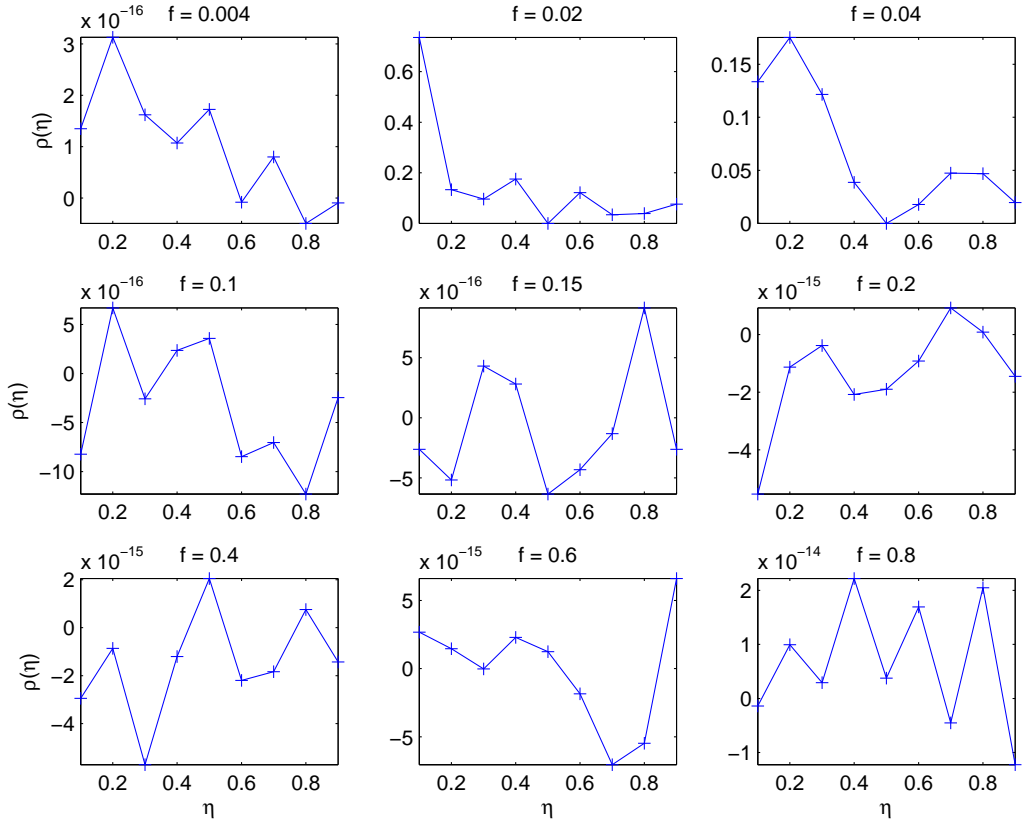}}
        \caption{Correlation coefficients $\rho$, cf. Eq.~(\ref{eq:rho}), of $a_f$ with $b_f$ (cf. Eqs.(\ref{eq:akr}), (\ref{eq:bkr})) against the $\eta$ parameter for a set of frequencies $f$ and a number of sampling time instants
       $M=500$ generated in such a way as to simulate five observing sessions of duration $100 \eta$ each containing $100$ equispaced data and covering a total fraction $\eta$ of the interval $[0, 500]$. Adjacent sessions are separated by a gap of 
        length $100 (1- \eta)$. Correlations are quite small except for $f=0.02$ when $\eta=0.1$.}
        \label{fig:corr3}
\end{figure*}

\begin{figure*}
        \resizebox{\hsize}{!}{\includegraphics{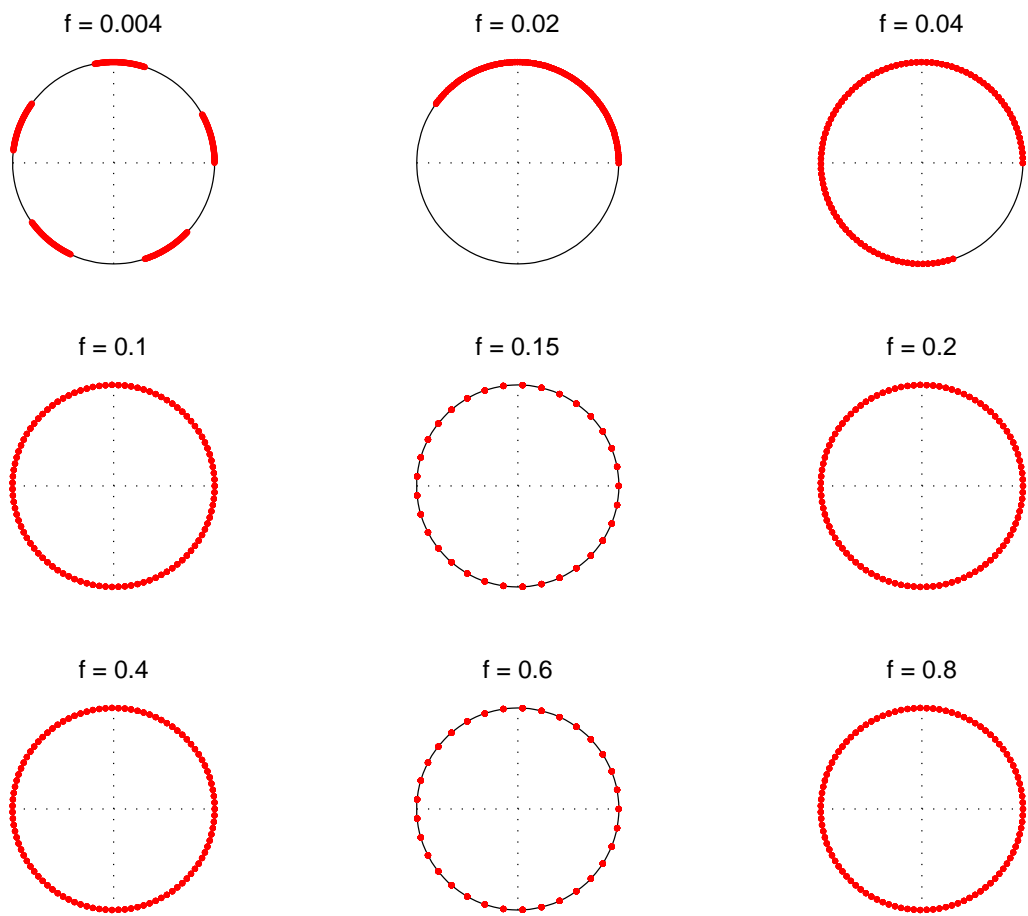}}
        \caption{Distribution of the angles $\alpha_j$ on the unit sphere for the set of frequencies $f$ and the sampling time instants as in Fig.~\ref{fig:corr3}  for the case $\eta=0.1$. Notice the distribution for
       $f=0.02$ that corresponds to the case when the correlation $\rho$ is high.}
        \label{fig:angles3}
\end{figure*}

To support these conclusions, in Figs.~\ref{fig:time1}-\ref{fig:angles3} the results of a few numerical simulations are presented. In particular, Figs.~\ref{fig:time1}-\ref{fig:time2} 
show the histograms of the  time instants corresponding to two sets of simulated sampling patterns ranging from regular to extremely irregular sampling. The reason for making such a choice is to verify that 
high values of $\rho$ are not linked to the degree of irregularity of the sampling.
For the first set, the time instants have been generated starting from a grid of time instants $\{ \underline{t}_j \}$ regularly spaced in the interval $[-1, 1]$ and then setting 
$t_j = \{ {\rm sign}[\underline{t}_j] ({\rm abs}[\underline{t}_j])^{\gamma} (M-1)+1 \}/2$. For the second set, the starting regular grid
is $\underline{t}_j \in [0, -1]$ and $t_j = \underline{t}_j^{\gamma} (M-1)$. Here, ${\rm sign}[.]$ is the {\it sign function}, ${\rm abs}[.]$ indicates {\it absolute value} and $\gamma$ is a positive real number. In both cases
$\gamma=1$ corresponds to an equispaced time grid. In the numerical experiment it has been assumed that $\sigma_n=1$ and several values of $\gamma$ have been tested. Figs.~\ref{fig:corr1}, \ref{fig:corr2} 
show the corresponding correlation coefficients. The values of $M=100$ and $M=1000$ have been taken as cases of a small and of a larger data set, respectively. In both cases,
 the median time sampling step for the different values of $\gamma$ lies approximately within the interval $(0.4, 1.1)$. A set of frequencies has been examined in the range $[0.1, 3.0]$. 
From these figures it is clear that, in spite of the extremely irregular sampling 
under examination, significant correlation between $a_f$ and $b_f$ happens only for the small data set. Even in this case, the correlation is weak ($ \le 0.2$). 
The fact that $\rho$ depends on  the distribution of the angle $\alpha_j$ is supported by Figs.~\ref{fig:angles1}, \ref{fig:angles2} where the distribution of the angles 
$\{ \alpha_j \}$ on the unit circle is shown for the case $M=100$ and $\gamma = 2.5$. It is evident that also with this limited number of data, the distribution of the $\alpha_j$ is approximately  uniform and symmetric.
For the case $M=1000$, the distribution (not shown here) is even more regular and indeed, as visible in Figs.~\ref{fig:corr1}, \ref{fig:corr2} 
the corresponding correlation coefficients are closer to zero.

Figures~\ref{fig:corr3}-\ref{fig:angles3} show the results concerning a few sampling patterns of more astronomical interest. In particular, for each value of $\eta$, chosen in the range $[0.1, 0.9]$,
$500$ sampling time instants $t_j$ have been generated according to
\begin{equation}
t_j = (j-1) +(100 \eta  - 99)/99 \times  {\rm mod}(j-1,100), \qquad j=1,2,\ldots, 500.
\end{equation}
In this way, five equispaced observing sessions of duration $100 \eta$ are simulated each containing $100$ equispaced data and covering a total fraction $\eta$ of the interval $[0, 500]$. Adjacent sessions are separated by a gap of length $100 (1- \eta)$. 
Fig.~\ref{fig:corr3} shows the correlation coefficients $\rho$ for a set of frequencies $f$ corresponding to different values of $\eta$. Again, most of them are small. Only for $\eta=0.1$ (i.e. very large gaps) and $f=0.02$, does $\rho$ 
attain the value $\approx 0.7$.
Fig.~\ref{fig:angles3} shows the distribution of the angles $\{ \alpha_j \}$ on the unit circle computed for $\eta=0.1$.
So significant a correlation is due to a sampling pattern of the type
given by Eq.~(\ref{eq:irrs}). A significant correlation for a  given frequency does not imply that the same holds for other frequencies.

\subsection{Analysis of an experimental time series}

\begin{figure*}
        \resizebox{\hsize}{!}{\includegraphics{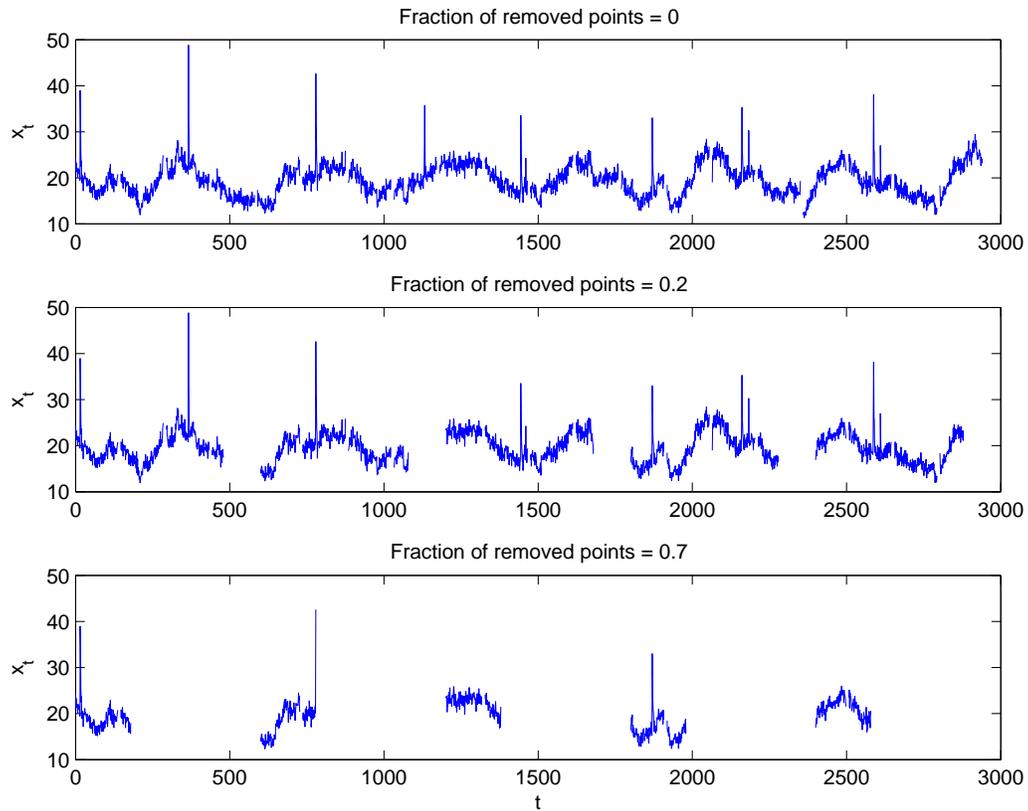}}
        \caption{Top panel: original optical light curve of the  transient low mass X-ray binaries \exo\. The time is in units of $\Delta t = 32 \rm{s}$; Central and bottom panels: the same light curve with $20\%$ and $70\%$ of the 
        data removed in such a way as to simulate $5$ different observing sessions with the same duration and spaced with gaps again with the same duration.}
        \label{fig:series}
\end{figure*}

\begin{figure*}
        \resizebox{\hsize}{!}{\includegraphics{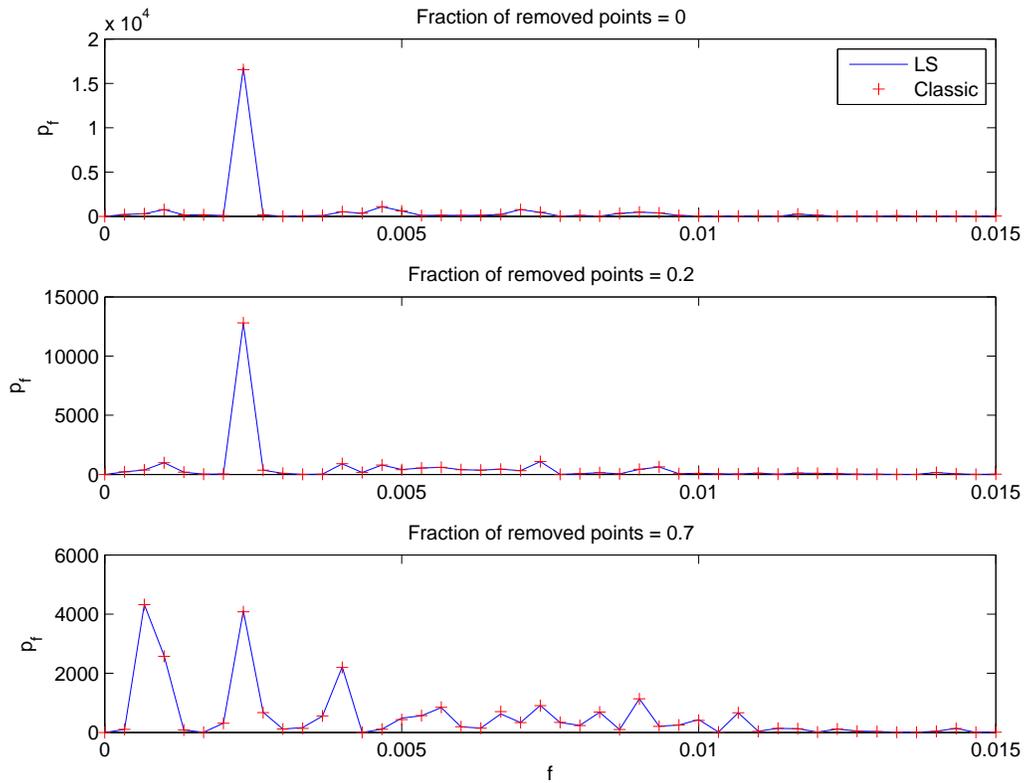}}
        \caption{Lomb-Scargle periodogram (LS) vs. the periodogram as given by Eq.~(\ref{eq:pf})  (here indicated as ``{\it classic}'') corresponding to the mean-subtracted time series in Fig.~\ref{fig:series}. 
        The frequency is in units of $1/\Delta t$ with $\Delta t$ the median sampling time step of the original light curve from which the time series have been obtained.}
        \label{fig:powers}
\end{figure*}

\begin{figure*}
        \resizebox{\hsize}{!}{\includegraphics{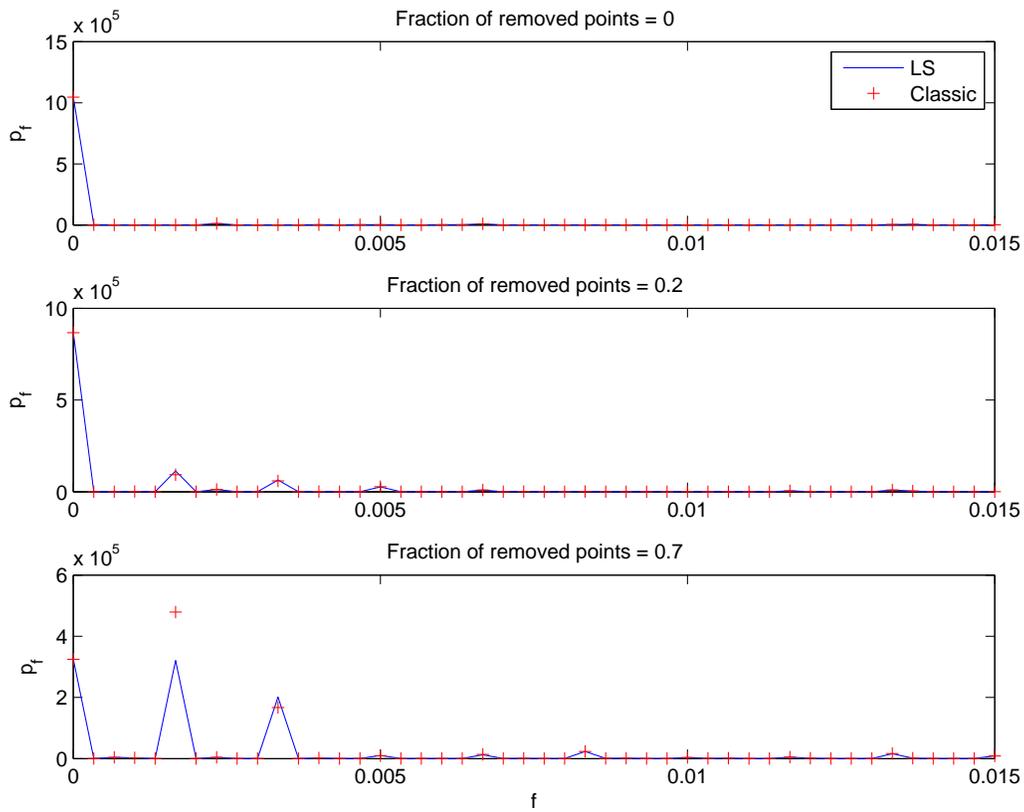}}
        \caption{Like in Fig.~\ref{fig:powers} but without the subtraction of the mean from the signal. Notice that, unlike for Fig.~\ref{fig:powers}, here the Lomb-Scargle periodogram (LS) is different from the  periodogram
as given by Eq.~(\ref{eq:pf}) (here indicated as ``{\it classic}'') .}
        \label{fig:powers1}
\end{figure*}

For demonstration purposes, we check what happens in the case of an experimental time series with periodic gaps. As explained above, this is a situation more favorable for a nonuniform distribution of the
angles $\alpha_j$.  In this regard, the LS periodogram and the version as given in Eq.~(\ref{eq:pf}) are compared in the case of the light curve of the low mass X-ray binary \exo\, a source 
which shows an orbital period of 3.82 hr \citep{par86}. This object shows 8.3 minute X-ray eclipses every orbital period,
irregular dipping activity (energy-dependent absorption) and type I X-ray bursts.  
\exo\ has been extensively studied with the X-ray observatory \xmm. In particular, it was observed on seven occasions during 
September--November 2003 for a total exposure time of 570~ks. For each observation, data were acquired simultaneously with all of the on-board instruments. Here, we present the light curve of the optical/UV monitor 
\citep{mas01} during 12-13 November 2003. The data were originally taken with a sampling of 500 ms. These data are
quite noisy and the light curve was rebinned to a sampling of 32 s to increase the signal-to-noise ratio per bin.
The sampling of this signal, shown in the top panel of Fig.~\ref{fig:series}, is regular but some
periodic gaps are present. Since these gaps are rather short, we have considered two other situations where larger periodic gaps are obtained by removing $20\%$ and $70\%$ of the data as shown in the central and bottom
panels of Fig.~\ref{fig:series}, respectively. In this way a time series with regular sampling and wider periodic gaps is obtained.
In spite of the presence of large gaps,  from Fig.~\ref{fig:powers} it clearly appears that the periodogram~(\ref{eq:pf}) and the LS periodogram when computed for the mean-subtracted signal, are quite similar. As is visible in Fig.~\ref{fig:powers1},
the same is not true without subtraction of the sampling mean. This is not surprising since, as shown above, the mean value introduces a spurious correlation between the $a_f$ and $b_f$ coefficients.

\section{Final remarks and conclusions} \label{sec:conclusions}

In this paper we have addressed the problems related to the spectral analysis of uneven time series. We have reexamined, with formalized arguments and some numerical experiments, the pros and cons of an even vs.
an uneven sampling.
\begin{enumerate}
\item A regular data sampling simplifies the analysis as well as the development of efficient algorithms. However, it permits one to retrieve only the frequencies characteristic of the signal that are smaller than the {\it Nyquist frequency}; 
\item An irregular sampling introduces some computational as well statistical problems but it permits one to retrieve information about frequencies even much greater than the {\it Nyquist frequency}; 
\item Although from the theoretical point of view techniques specific to the spectral analysis of uneven sampled 
signals such as the Lomb-Scargle periodogram  could be of some interest, their effectiveness in practical astronomical applications is limited. Indeed, approximated but simpler techniques are able to provide similar results and are easier 
to use and to modify to deal with situations different from those under which the original LS periodogram has been developed.
\end{enumerate}
Before concluding, it is necessary to stress that often in Astronomy the spectral analysis can be safely used only as a test to check whether a time series contains a signal of interest or is constituted only of noise. 
For example, for both the regular and the irregular sampling, the periodogram cannot provide a reliable statistical characterization of a {\it red noise} signal if the experimental time series spans an interval shorter than the time scale of the signal itself.
Moreover, in the presence of an irregular sampling and independently of the technique used, the periodogram cannot be used to identify the frequencies of a periodic signal because of the ``{\it interference}'' between the true peaks and those 
due to sampling \citep{dee75}. In this case, other techniques are necessary \citep[e.g. ][]{rob87, fos95, bou07}.

Some software code and data can be made available upon request.

\clearpage




\begin{thebibliography}{}
\bibitem[Bj\"orck (1996)]{bjo96} Bj\"orck A. 1996, Numerical Methods for Least Squares Problems (Philadelphia: SIAM)
\bibitem[Bourguignon et al. (2007)]{bou07} Bourguignon, S., Carfantan, H., \& B\"ohm, T. 2007, A\&A, 462, 379
\bibitem[Chu (2008)]{chu08} Chu, E. 2008, Discrete and Continuous Fourier Transform (Boca Raton: Chapman \& Hall{\textbackslash}CRC)
\bibitem[Cumming et al. (1999)]{cum99} Cumming, A., Marcy, G.W., \& Butler, R.P. 1999, ApJ, 526, 890
\bibitem[Deeming (1975)]{dee75} Deeming, T.J. 1975 Ap\&SS, 36, 137D
\bibitem[Eyer \& Bartholdi (1999)]{eye99} Eyer, L., \& Bartholdi, P. 1999, A\&A Suppl. Ser., 135, 1
\bibitem[Foster (1995)]{fos95} Foster, G. 1995, AJ, 109, 1889
\bibitem[Gottlieb et al. (1975)]{got75} Gottlieb, E.W., Write, E.L., \& Liller, W. 1975, ApJL, 195, L33
\bibitem[Keiner et al. (2008)]{kei08} Keiner, J., Kunis, S., \& Potts, D. 2008, ACM Transactions on Mathematical Software, 5, 1
\bibitem[Koen (2006)]{koe06} Koen, C. 2006, MNRAS, 371, 1390
\bibitem[Lomb (1976)]{lom76} Lomb, N.R. 1976, Ap\&SS, 39, 447
\bibitem[Mason et al. (2001)]{mas01} Mason, K.O. et al. 2001, A\&A, 365, L36
\bibitem[Parmar et al. (1986)]{par86} Parmar, A.N. et al. 1986, ApJ, 308, 199
\bibitem[Press et al. (2007)]{pre07} Press, W.H., Teukolsky, S.A., Vetterling, W.T., \& Flannery, B.P. 2007, Numerical Recipes
(Cambridge: Cambridge University Press)
\bibitem[Reegen (2007)]{ree07} Reegen, P. 2007, A\&A, 467, 1353 
\bibitem[Roberts et al. (1987)]{rob87} Roberts D.H., Lehar J., \&  Dreher J.W. 1987 ApJ, 93, 968 
\bibitem[Scargle (1982)]{sca82} Scargle, J.D. 1982, ApJ, 263, 835  
\bibitem[Stoica et al. (2009)]{sto09} Stoica, P., Li, J., \& He, H. 2009, IEEE Transaction of Signal Processing, 57, 843
\bibitem [Vio et al. (2010)]{vio10} Vio, R., Andreani, P., \& Biggs, A. 2010, A\&A, 519, A85
\bibitem[Zechmeister \& K\"urster (2009)]{zec09} Zechmeister, M., \& K\"urster, M. 2009, A\&A, 496, 577 
\end{thebibliography}
\end{document}